\documentclass[galaxies,article,submit,moreauthors,pdftex,10pt,a4paper]{mdpi2} 

\firstpage{1} 
\articlenumber{x}
\doinum{10.3390/------}
\pubvolume{xx}
\pubyear{2016}
\copyrightyear{2016}
\history{Received: date; Accepted: date; Published: date}
\pdfoutput=1

\Title{Observing---and Imaging---Active Galactic Nuclei with the Event
  Horizon Telescope}

\Author{Vincent L.\ Fish $^{1}$*, Kazunori Akiyama $^{1,2}$, Katherine
  L.\ Bouman $^{3}$, Andrew A.\ Chael $^{4}$, Michael D.\ Johnson
  $^{4}$, Sheperd S.\ Doeleman $^{4}$, Lindy Blackburn $^{4}$, John
  F.\ C.\ Wardle $^{5}$, William T.\ Freeman $^{3,6}$, and the Event
  Horizon Telescope Collaboration}
\AuthorNames{Vincent Fish, Kazunori Akiyama, Katherine Bouman, Andrew
  Chael, Michael Johnson, Sheperd Doeleman, Lindy Blackburn, John
  Wardle, William Freeman, EHT Collaboration}

\address{%
$^{1}$ \quad Massachusetts Institute of Technology, Haystack Observatory\\
$^{2}$ \quad Japan Society for the Promotion of Science, Postdoctoral
  Fellow for Research Abroad \\
$^{3}$ \quad Massachusetts Institute of Technology, Computer Science
  and Artificial Intelligence Laboratory \\
$^{4}$ \quad Harvard-Smithsonian Center for Astrophysics\\
$^{5}$ \quad Brandeis University, Physics Department \\
$^{6}$ \quad Google}

\corres{Correspondence: vfish@haystack.mit.edu; Tel.: +1-781-981-5400}

\simplesumm{The Event Horizon Telescope (EHT) observes black holes
  using a global array of telescopes.  This work summarizes recent
  results from the EHT, with emphasis on new ways to make accurate
  images from the data.  These techniques are applicable to data from
  other arrays too.}

\abstract{Originally developed to image the shadow region of the
  central black hole in Sagittarius A* and in the nearby galaxy M87,
  the Event Horizon Telescope (EHT) provides deep, very high angular
  resolution data on other AGN sources too.  The challenges of working
  with EHT data have spurred the development of new image
  reconstruction algorithms.  This work briefly reviews the status of
  the EHT and its utility for observing AGN sources, with emphasis on
  novel imaging techniques that offer the promise of better
  reconstructions at 1.3 mm and other wavelengths.}


\conferencetitle{Blazars through Sharp Multi-wavelength Eyes}

\begin{document}

\section{The Event Horizon Telescope}

\subsection{Overview}

The Event Horizon Telescope (EHT) is a very long baseline
interferometry (VLBI) array currently operating at a wavelength of
1.3~mm, with extension to 0.87~mm to come.  A primary goal of the EHT
is to observe and image supermassive black holes with enough angular
resolution to study the innermost accretion and outflow region at a
few gravitational radii from the black hole \cite{doeleman2009}.

The EHT consists of millimeter/submillimeter telescopes scattered
across the globe, including the Submillimeter Array and James Clerk
Maxwell Telescope in Hawaii, the Submillimeter Telescope on
Mt.\ Graham in Arizona, the Large Millimeter Telescope in Mexico, the
Atacama Pathfinder Experiment in Chile, the South Pole Telescope, the
Institut de radioastronomie millim\'{e}trique (IRAM) 30-m telescope on
Pico Veleta in Spain, and the IRAM Plateau de Bure Interferometer in
France (currently being upgraded as part of the Northern Extended
Millimeter Array project).  A beamformer has been built for the
Atacama Large Millimeter/submillimeter Array (ALMA) to allow it to
operate as a very sensitive single aperture for VLBI.  Several other
observatories have participated in previous EHT observations, and
additional observatories are on track to join the array in upcoming
years.

The angular resolution of the EHT is especially well matched to the
scale of emission in the black holes in the center of the Milky Way
(known as Sagittarius~A* or Sgr A*) and the giant elliptical galaxy
M87.  General relativity predicts that if a black hole is surrounded
by optically thin emission, the black hole will cast an approximately
circular shadow with a diameter of around 10 gravitational radii (1
$r_g = GMc^{-2}$), which corresponds to about 50~$\mu$as in Sgr~A* and
40~$\mu$as in M87.  Fringe spacings of EHT baselines at 1.3~mm span a
range of approximately 25 to 300~$\mu$as, providing both the
resolution needed to image the shadow region and shorter spacings to
be sensitive to the accretion and outflow region out to a few tens of
$r_g$.  The very high angular resolution of the EHT is also useful for
studying compact structures in other AGN sources.  For instance, EHT
observations of the quasars 1924$-$292 and 3C~279 have demonstrated
that the small structures seen at 1.3~mm have lower brightness
temperatures than seen at centimeter wavelengths
\cite{lu2012,lu2013,wagner2015}.

\subsection{Key Science Results on Sgr~A* and M87}

Sgr~A* has been routinely detected on long EHT baselines since 2007,
and M87 since 2009.  Although early EHT experiments have not had the
sensitivity or baseline coverage to directly reconstruct images of
these targets or other sources, a large number of important scientific
results have been derived from the visibility data obtained to date.
A very brief summary of these results appears below, with full details
to be found in the referenced papers.

\begin{itemize}[leftmargin=*,labelsep=4mm]
\item The 1.3~mm emission regions in Sgr~A* and M87 have an effective
  size of only a few $r_g$ \cite{doeleman2008,doeleman2012}.
\item The small sizes of Sgr~A* and M87 combined with their relatively
  weak infrared emission make a very strong case that these sources
  have an event horizon and are therefore black holes
  \cite{broderick2009,broderick2015}.
\item The millimeter-wavelength emission region in Sgr~A* remains
  compact when the source flares \cite{fish2011}.
\item Sgr~A* is asymmetric on angular scales comparable to the shadow
  diameter.  This asymmetry is persistent and intrinsic to the source,
  rather than arising from the foreground scattering screen
  \cite{fish2016}.
\item The magnetic field in the inner accretion flow around Sgr~A* is
  partially ordered.  The polarized millimeter-wavelength flux arises
  from the same region as the total intensity during quiescence, but
  polarized variability can occur in regions that are offset from the
  main emission region \cite{johnson2015}.
\item The data strongly constrain specific model classes for Sgr~A*
  and M87.  In most cases, these models find that the black hole spin
  vector is inclined significantly with respect to the line of sight
  \cite{broderick2009b,moscibrodzka2009,dexter2010,chan2015}.
\item The consistency of the size of the millimeter-wavelength
  emission in M87 before and during a very high energy flare favors
  models in which $\gamma$ rays originate in an extended region
  \cite{akiyama2015}.
\end{itemize}

\subsection{Observing AGN Sources with the EHT}

The EHT is uniquely capable of probing the innermost regions of other
AGN sources at high angular resolution.  The EHT penetrates more
deeply into the synchrotron photosphere of jets than any other,
lower-frequency VLBI array.  With an angular resolution of better than
25~$\mu$as, the EHT provides superior resolving power on AGN sources,
which is exceeded only by RadioAstron---and even then only at greatest
elongation, at its highest frequency (22~GHz, a factor of 10 smaller
than the EHT), and in a single direction.  The combination of high
angular resolution and high frequency allows the EHT to measure jet
widths much closer to the launch point than lower-frequency
observations, enabling detailed studies of the inner collimation
region.  As a high-frequency, wide-bandwidth observing array, the EHT
will also be able to detect and measure regions of high rotation
measure in jet sources \cite{vertatschitsch2015}.  With the expansion
of the EHT to a total data rate of 64~Gb\,s$^{-1}$ split across two
noncontiguous sidebands, the spanned frequency will be large enough to
study sources with high rotation measures, such as 3C~84
\cite{plambeck2014}.

\begin{figure}[H]
\centering
\includegraphics[width=6cm]{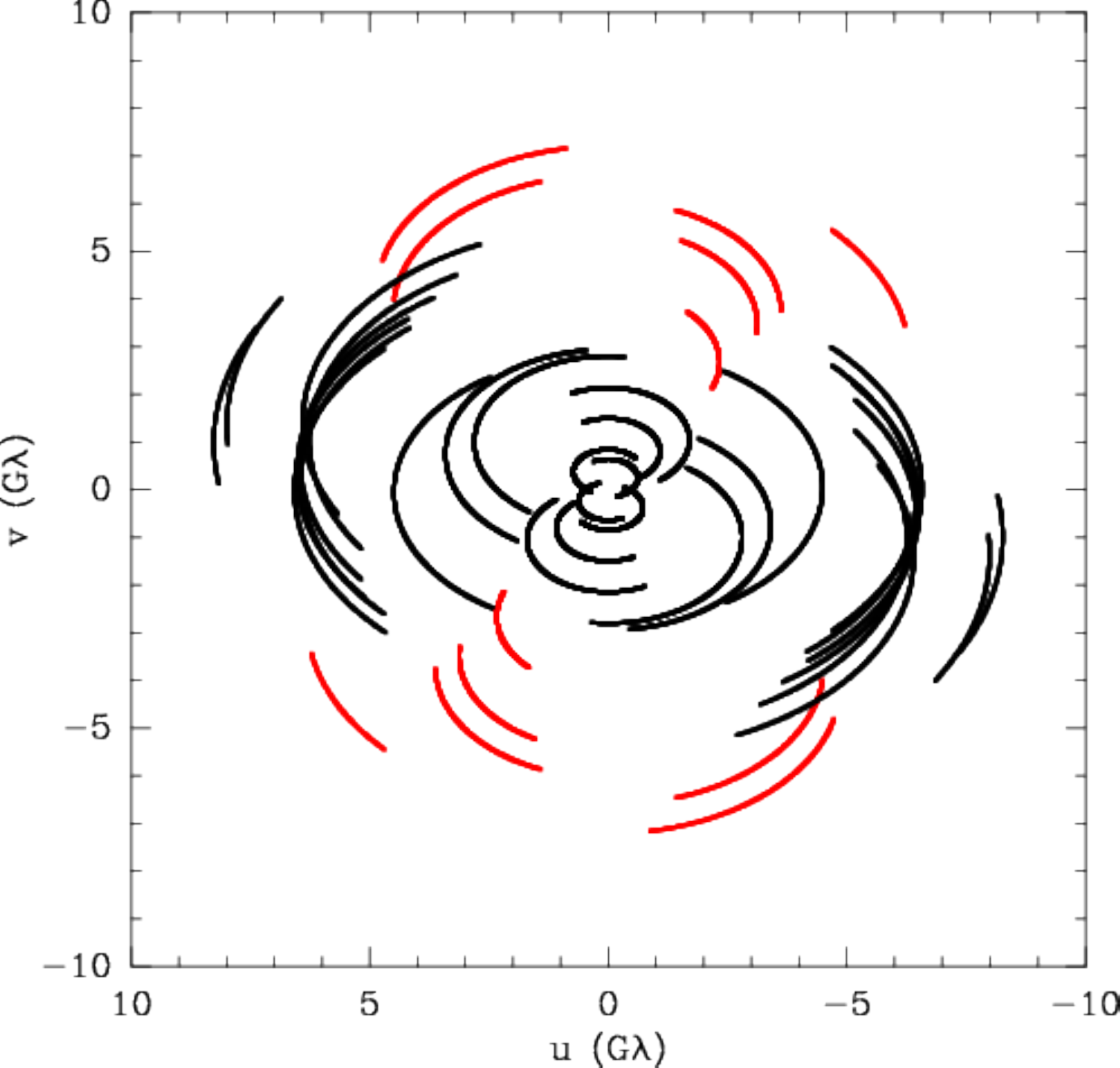}
\includegraphics[width=6cm]{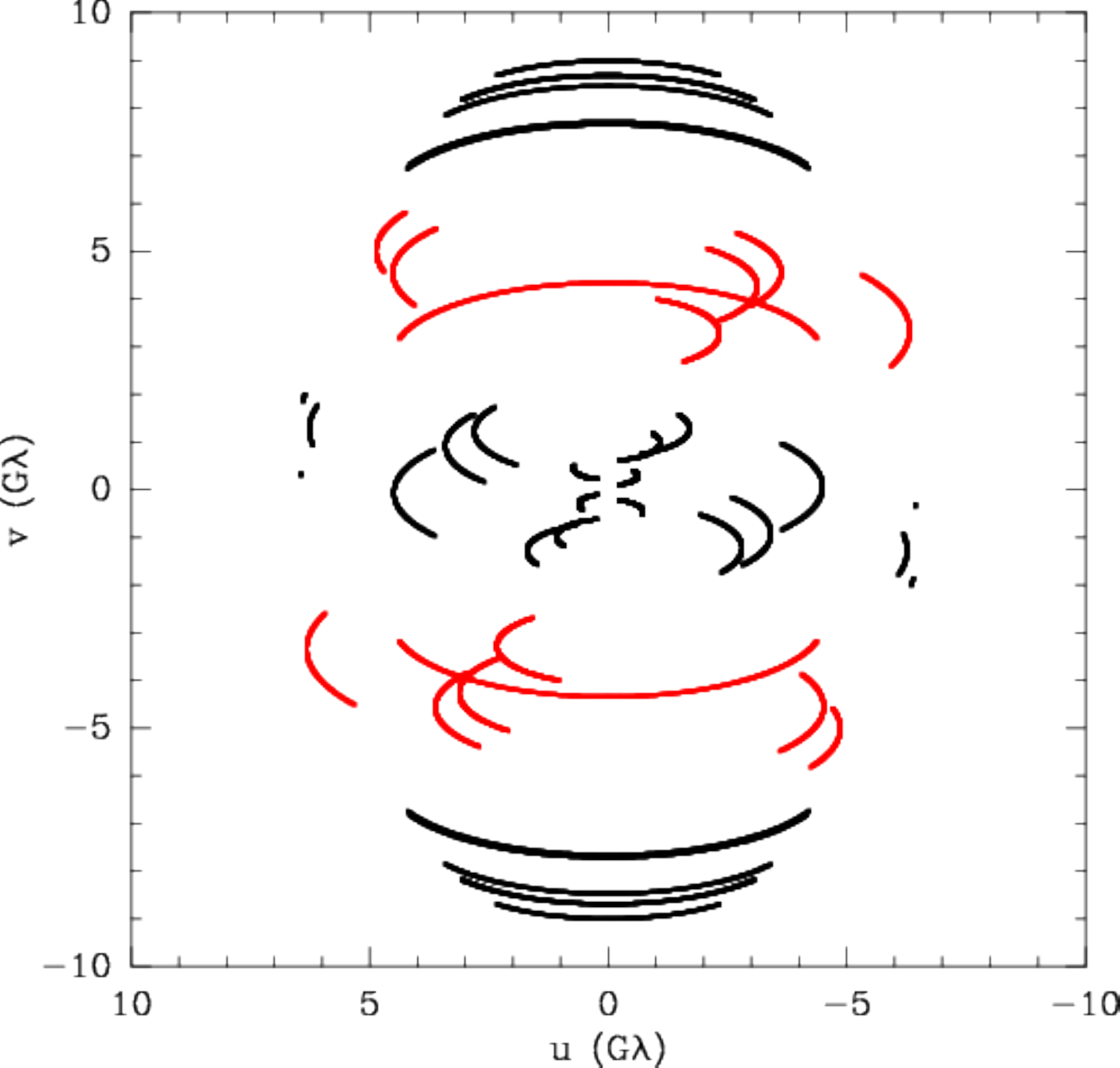}
\caption{EHT 1.3~mm baseline coverage for a source at $+40^\circ$
  declination (\textbf{a}) and $-20^\circ$ (\textbf{b}).  Baselines to
  ALMA are shown in red.  ALMA baselines extend north-south coverage
  and are especially important for southern sources, where they fill
  in a large gap in the $(u,v)$ plane between the intra--northern
  hemisphere baselines and the very long baselines to the South Pole
  Telescope.
  \label{uvtracks}}
\end{figure}

The range of scientific projects that can be undertaken with the EHT
is very large \cite{fishapp,tilanusapp}.  The ALMA Phasing Project has
created a beamformer for ALMA, which enhances the sensitivity of the
EHT array and thus greatly extends the range of sources that can be
observed with the EHT.  As part of the ALMA Cycle~4 Call for
Proposals, the ALMA phasing system was for the first time offered for
open-access VLBI observing in conjunction with the EHT at 1.3~mm and
the Global mm-VLBI Array (GMVA) at 3.5~mm.  Inclusion of ALMA also
extends the north-south resolution of these VLBI arrays
(Figure~\ref{uvtracks}).  Thanks to the increases in sensitivity and
baseline coverage, VLBI arrays that make use of phased ALMA will be
able to make better images than have heretofore been possible.

\section{Imaging}

Reconstructing an image of the observed target is a critical part of
VLBI observing, no matter which array is used or which frequency is
observed.  VLBI arrays generally sample the $(u,v)$ plane only
sparsely, leaving many Fourier components of the sky image unsampled.
Image reconstruction methods make use of the limited data (e.g.,
measured visibilities) along with externally imposed constraints
(e.g., positive definiteness in Stokes~I images) to select one image
out of the many that are consistent with the data.

\subsection{CLEAN}

The standard image reconstruction algorithm for radio interferometer
data is CLEAN \citep{hogbom1974}.  Visibilities are gridded and
Fourier transformed to produce a dirty map that is contaminated by the
sidelobe structure of the dirty beam.  To remove the sidelobe
structure, CLEAN runs an iterative loop in which the dirty beam is
recentered at the brightest pixel in the dirty map and a fraction of
the brightness of that pixel, along with its sidelobe contribution to
other pixels, is subtracted.  This loop is run until an end condition
is reached, typically determined by the number of loop iterations, the
noise level in the residual map, or the effective dynamic range of the
image.  The CLEAN components are then convolved with a restoring beam
and added back to the residual map.

In the many years since CLEAN was introduced, numerous variants of the
algorithm have been implemented to optimize for speed, suppress
artifacts, handle extended sources, image wide fields, and deal with
spectral structure across wide bandwidths.  All of these variations
use deconvolution to remove the dirty beam structure from the dirty
map.  CLEAN variants have become very popular in radio interferometry,
due both to their success in producing reasonable images from most
radio arrays and to their implementation in common radio
interferometry data reduction packages, such as AIPS, CASA, and
Miriad.

\subsection{Imaging with the EHT}

CLEAN has been successful in producing interferometric images of
adequate quality at lower frequencies.  However, it is less well
suited to reconstructing images from the EHT for several reasons.

First, the $(u,v)$ coverage of the EHT is sparse due to the limited
number of observatory sites and their distribution on the Earth.  The
resulting dirty beam of the EHT has large sidelobes, which can lead to
large artifacts in images reconstructed by CLEAN.

Second, the fundamental VLBI observable used by CLEAN is the
(calibrated) complex visibility on each baseline in each Stokes
parameter.  Rapidly varying tropospheric delays preclude
phase-calibration of the data via the standard technique of nodding
between the target source and a nearby calibrator.  The EHT
compensates for this by using robust VLBI quantities where possible.
The robust Stokes I phase observable is the closure phase, which is
the directed sum of visibility phases around a closed triangle of
stations\footnote{or equivalently, the phase of the bispectrum, which
  is the product of the three complex visibilities}
\cite{jennison1958,rogers1974}.  When possible, it is also preferable
to use closure amplitudes---ratios of visibility amplitudes among
quadrangles of stations---that are less sensitive to gain fluctuations
at each antenna.  For linear polarization, it is most natural to use
complex polarimetric ratios such as $RL/LL$, where $L$ and $R$ refer
to the left and right circular polarizations, which are robust against
both atmospheric phase variations and gain fluctuations
\cite{roberts1994}.  Hybrid mapping can use CLEAN to incorporate
closure information into station-based complex gains, but it is
preferable to work with the robust observables directly.

Third, the effective angular resolution provided by CLEAN is coarser
than that provided by, e.g., the Maximum Entropy Method (MEM)
\cite{cornwell1985}.  The diameters of the predicted black hole
shadows in Sgr~A* and M87 are only approximately twice the fringe
spacing of the longest baselines at 1.3~mm.  This is sufficient for
producing an image of the shadow region to examine the morphology and
size of the shadow, but angular resolution finer than the fringe
spacing is desirable in order to produce a clearer image that is more
suitable for testing general relativity
\cite{doeleman2009,johannsen2010,broderick2014,goddi2016}.  While it
is possible to create slightly superresolved images with CLEAN, MEM is
better at squeezing more angular resolution out of a dataset
\cite{narayan1986}.

Fourth, it is more natural to encode prior knowledge about the image
into techniques that fit to the visibilities directly rather than
CLEAN, which is a greedy method to deconvolve the dirty beam from a
dirty image.  Even before taking any data on a source, very strong
statements can be made about the range of images that are possible.
For instance, pixels in a typical Stokes I image should be
nonnegative, including many pixels that have no source flux in them at
all.  Typical images may have regions that vary smoothly and regions
with sharp gradients, but flux is not distributed randomly among image
pixels.  Some constraints can be used with deconvolution methods, but
it is simpler to add these as regularizers in methods that solve for
an image in the data domain.

\subsection{Lessons from Optical Interferometry}

The imaging challenges faced by the EHT bear a strong resemblance to
those confronted by the optical interferometry (OI) community
\cite{thiebaut2013}.  Optical interferometers typically consist of a
small number of telescopes, limiting the available $(u,v)$ coverage.
Atmospheric coherence times are orders of magnitude shorter at
near-infrared and optical wavelengths, necessitating the use of
closure phases or bispectra.  Power spectra, rather than visibility
amplitudes, are fundamental observables.  In addition, some OI imaging
targets, such as stellar disks, naturally present strong priors for
the form of the reconstructed image.

Recognizing the inadequacy of contemporary image reconstruction
techniques for OI data, the OI community has organized a biennial
imaging contest
\cite{contest2004,contest2006,contest2008,contest2010,contest2012,contest2014}.
Entries have spanned a wide range from adaptations of standard radio
interferometric imaging techniques to completely new algorithms.
Entries have included both deconvolution methods based on CLEAN and
forward-imaging methods in which reconstructed images are evaluated in
the data domain, the optimal image being the one that maximizes the
likelihood of the data with a regularization penalty.  In other words,
the optimal image minimizes the sum of $\chi^2$ and one or more
additional terms that quantify prior expectations of a
well-reconstructed image (e.g., sparseness or smoothness).
Forward-imaging methods usually produces images that are judged to be
closer to the simulated data presented in these imaging challenges.

Image reconstruction tests have demonstrated that the same
forward-imaging algorithms developed by the OI community are superior
to deconvolution methods for simulated EHT data \cite{lu2014}.  Two OI
imagers were tested: the BiSpectrum Maximum Entropy Method (BSMEM) and
SQUEEZE \cite{buscher1994,baron2010}.  Compared against multiscale
CLEAN, both BSMEM and SQUEEZE produced higher-fidelity images with
higher effective angular resolution \cite{cornwell2008}.

\subsection{New Imaging Algorithms for the EHT}

The success of modern image reconstruction methods has encouraged a
group within the EHT to investigate new algorithms that work with
robust VLBI observables.  Three of these methods are briefly described
in this subsection.

\subsubsection{Polarimetric MEM}

Maximum entropy principles, which favor sparsity, apply equally well
to polarimetric imaging.  Early tests demonstrated that MEM is
suitable for providing superresolution or for reconstructing extended
emission structures \cite{holdaway1990}.

\begin{figure}[H]
\centering
\includegraphics[width=9cm]{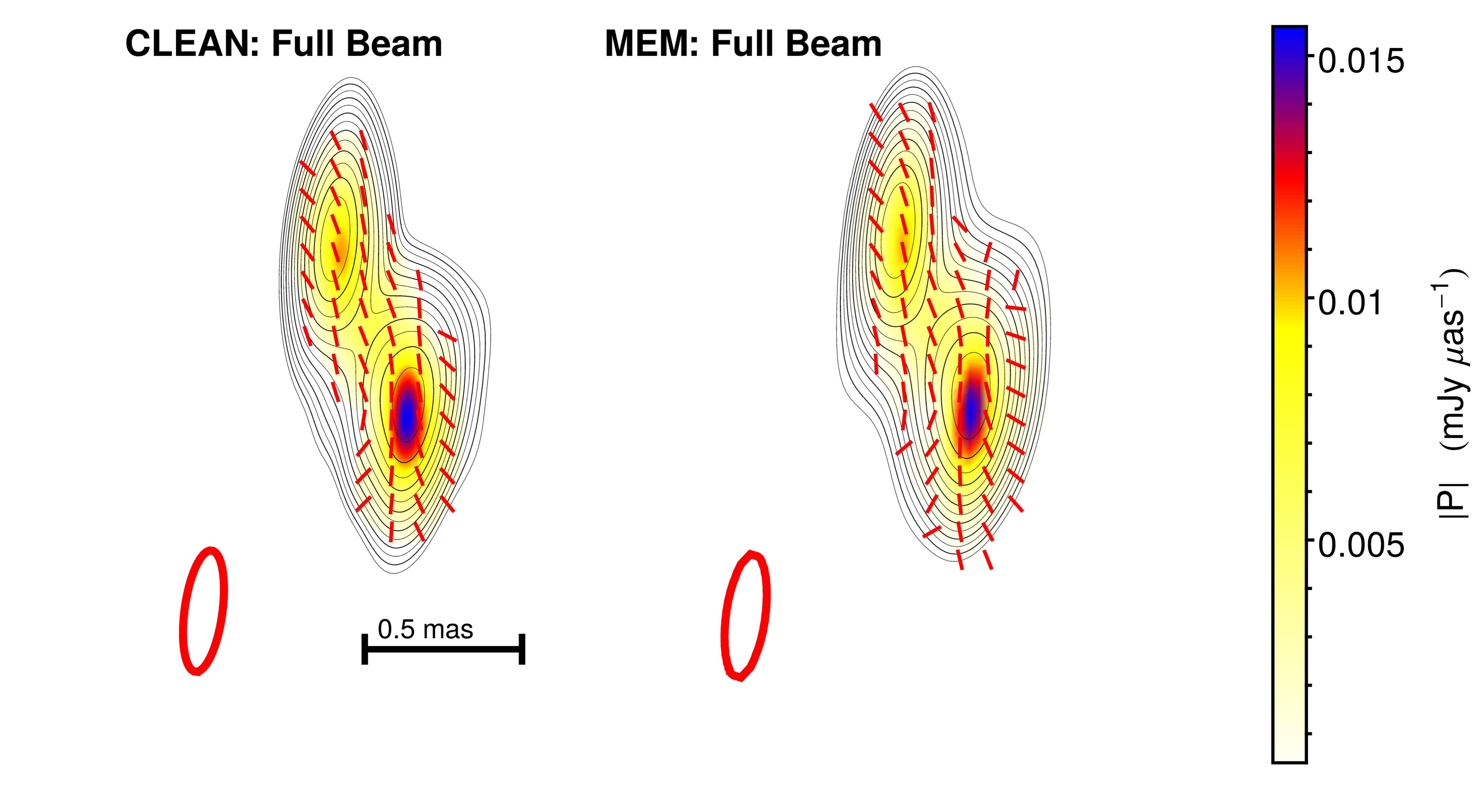}
\includegraphics[width=\textwidth]{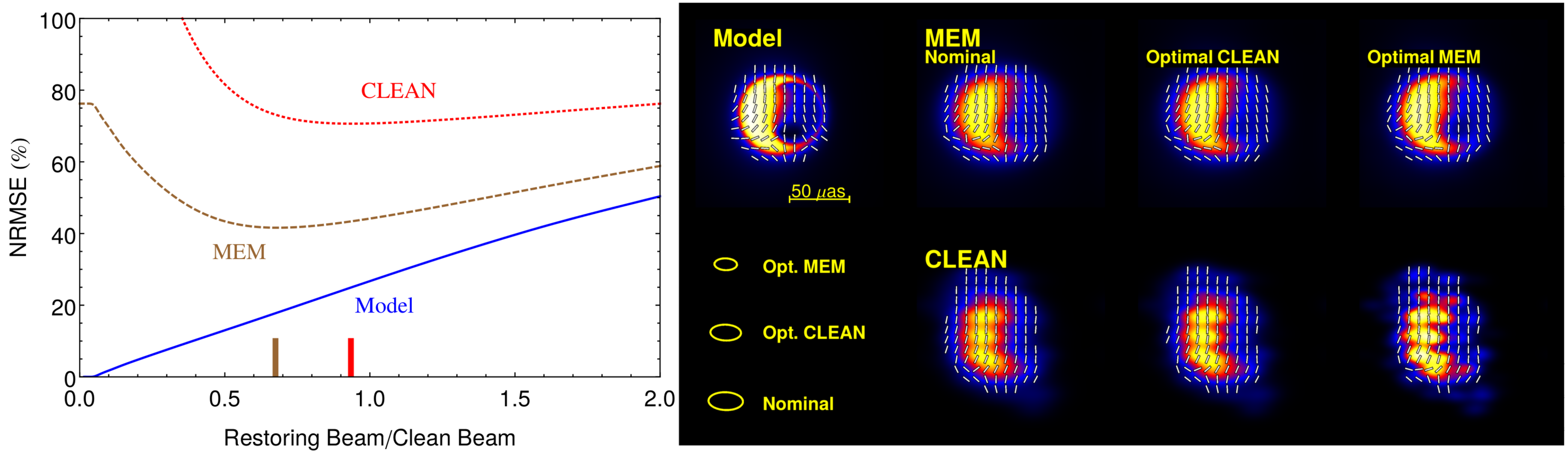}
\caption{Validation and potential of polarimetric MEM imaging
  \cite{chael2016}.  (\textbf{a}) Full-polarimetric images of 3C~279
  from 7~mm VLBA data using CLEAN and MEM are nearly identical when
  the MEM image is convolved with the CLEAN beam. (\textbf{b})
  Compared against a model image of Sgr~A* at 1.3~mm, the normalized
  root-mean-square error of a polarimetric MEM reconstruction achieves
  its minimum at finer angular resolution than does CLEAN.  MEM and
  CLEAN images reconstructed at the optimum beam sizes for each
  technique and the nominal beam size demonstrate that polarimetric
  MEM achieves superior resolution and image fidelity.
  \label{polmem}}
\end{figure}

Polarimetric MEM imaging has recently been extended to handle robust
VLBI observables, reconstructing Stokes I from bispectral data and
Stokes Q and U from polarimetric ratios \cite{chael2016}.  A variety
of different regularizers have been explored beyond traditional
total-intensity and polarimetric entropy terms, including regularizers
that favor a prior image and reconstructions using $\ell_1$ and
$\ell_2$ norms.  At reasonably high signal-to-noise, the reconstructed
image is not very sensitive to the choice of regularizer in either
total intensity or polarizartion, and the method has been shown to
produce good reconstructions even with 10\% systematic uncertainties
in visibility amplitudes.  This technique has been validated on
polarimetric 7~mm and 3~mm Very Long Baseline Array (VLBA) data, and
it shows great promise for imaging the polarized structure in the
inner accretion and outflow regions of Sgr~A* and M87
(Figure~\ref{polmem}).

\subsubsection{Bi-Spectrum Sparse Modeling (BSSpM)}

Compressed sensing techniques show promise for providing
superresolution of sparse, compact structures.  EHT simulations have
demonstrated that compressed sensing using the Least Absolute
Shrinkage and Selection Operator (LASSO), a regularizer based on the
$\ell_1$ norm, can reconstruct the shadow region of M87 quite well
using visibility data \cite{honma2014}.  BSSpM extends this method to
work with closure phases rather than visibility phases
\cite{akiyama2016}.

\begin{figure}[H]
\centering
\includegraphics[width=\textwidth]{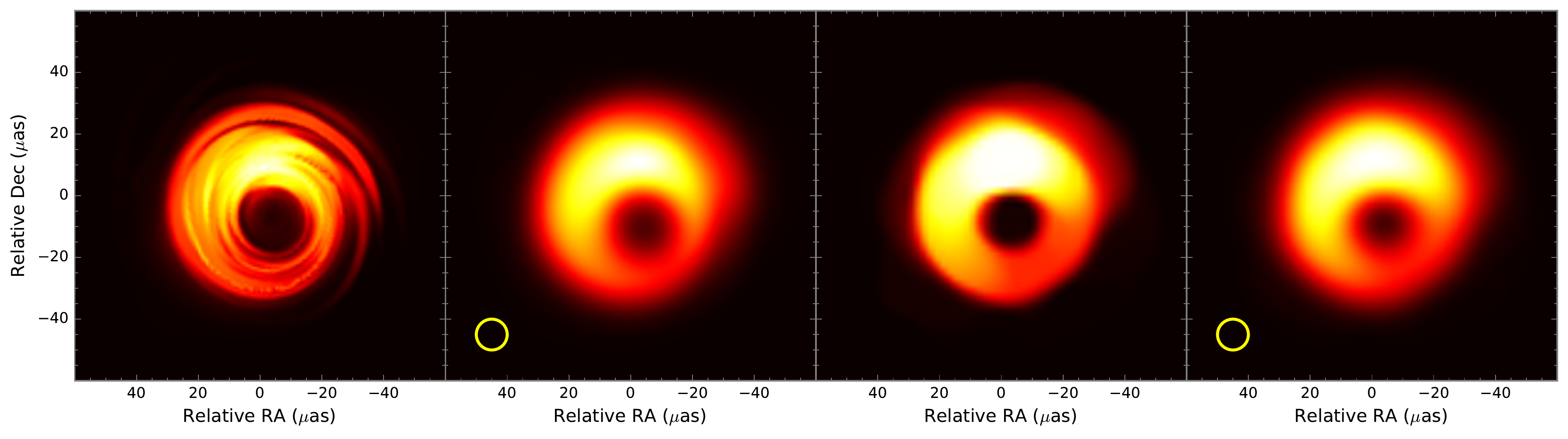}
\caption{BSSpM reconstruction of a disk-jet model of M87 at
  1.3~mm. (\textbf{a}) The original model \cite{dexter2012}.
  (\textbf{b}) The model convolved with a 10~$\mu$as Gaussian.
  (\textbf{c}) The BSSpM reconstruction of the image from simulated
  data \cite{akiyama2016}.  The fringe spacing of the longest baseline
  is approximately 25~$\mu$as.  (\textbf{d}) The BSSpM reconstruction
  convolved with a 10~$\mu$as Gaussian.  BSSpM successfully recovers
  features of the model at a superresolution of 40\% of the fringe
  spacing.
  \label{bsspm}}
\end{figure}

BSSpM also adds a second regularizer term for Total Variation (TV),
which favors sparsity in the gradient of the image.  Since TV
regularization is edge-preserving, it is well suited to images with
sharp boundaries between emitting and non-emitting regions, such as
the inner edge of the black hole shadow or edge-brightened jets.
Tests on simulated EHT data demonstrate that BSSpM has the potential
to superresolve the black hole shadow region (Figure~\ref{bsspm}).

\subsubsection{Continuous High-resolution Image Reconstruction using
  Patch priors (CHIRP)}

Since the problem of image reconstruction from sparse frequency
measurements is ill-posed, we always have to inject some sort of
information about what images look like, in the form of a regularizer
or prior, in order to decide on a reconstructed image.  However, it is
not clear what this injected information should be, since black
holes have not been imaged before.  In addition, prior assumptions
about what constitutes a good image may introduce biases into
reconstructions.  An alternative to hand-designing regularizers is to
design data-driven priors.  This provides the flexibility to easily
encode different image assumptions into the imaging process by
training the algorithm on different kinds of images.

CHIRP is an interferometric image reconstruction method that breaks
images up into patches of pixes and models the structure of building
blocks of different types of images \cite{bouman2016}.  A key insight
is that not all arrangements of pixel brightnesses within a patch of a
natural image are equally likely.  For instance, natural images often
contain some patches that are nearly uniform in intensity and others
with a sharp gradient, but patches whose pixel intensities are random
and uncorrelated are uncommon.  Algorithms can be trained on large
datasets of natural, celestial, or synthetic black hole images to
determine the likelihood that any particular patch occurs.  CHIRP
solves for the image that is the best fit to the bispectral data,
using the expected patch log likelihood as a regularizer.

\begin{figure}[H]
\centering
\includegraphics[width=10cm]{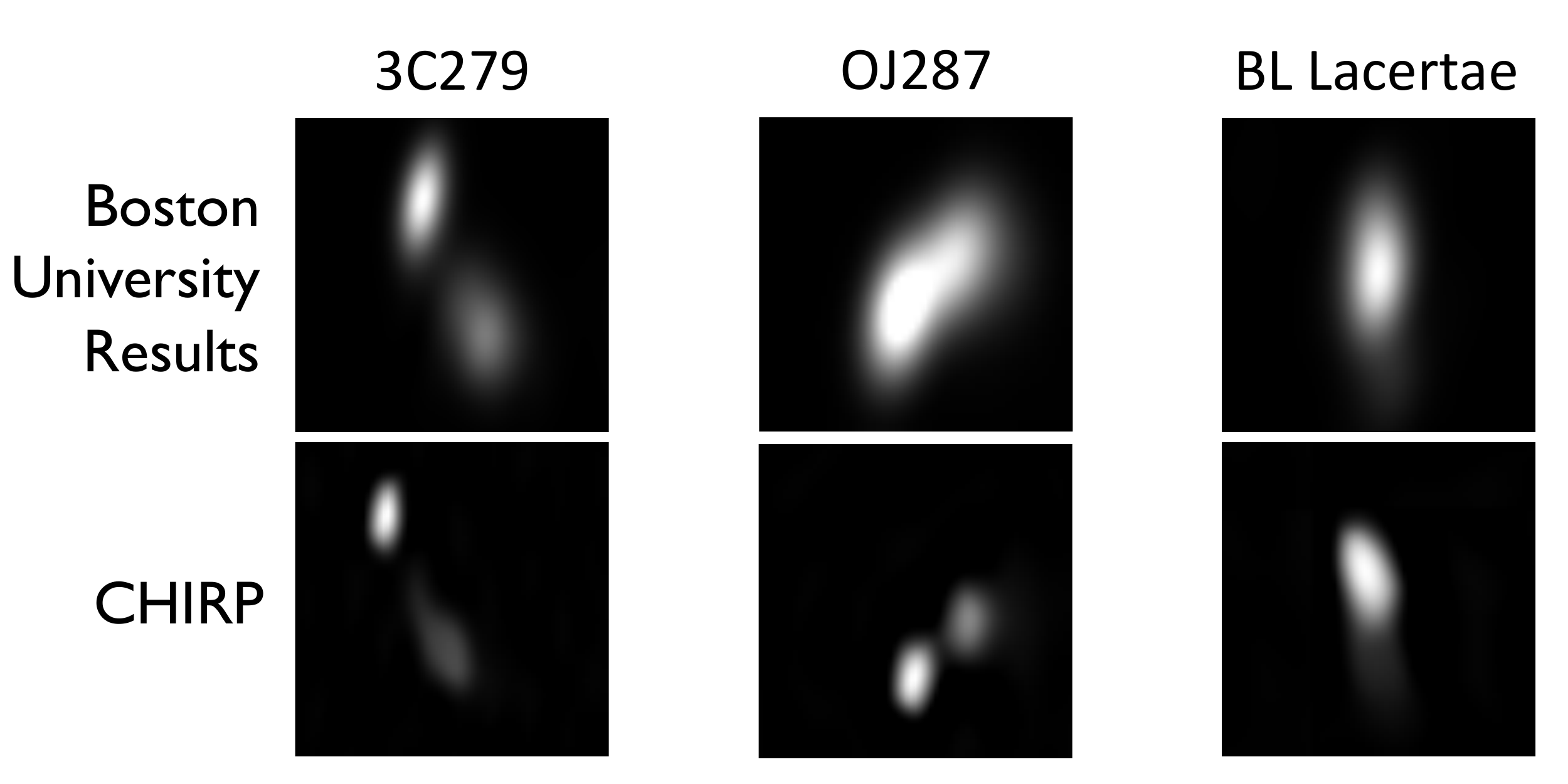}
\includegraphics[width=10cm]{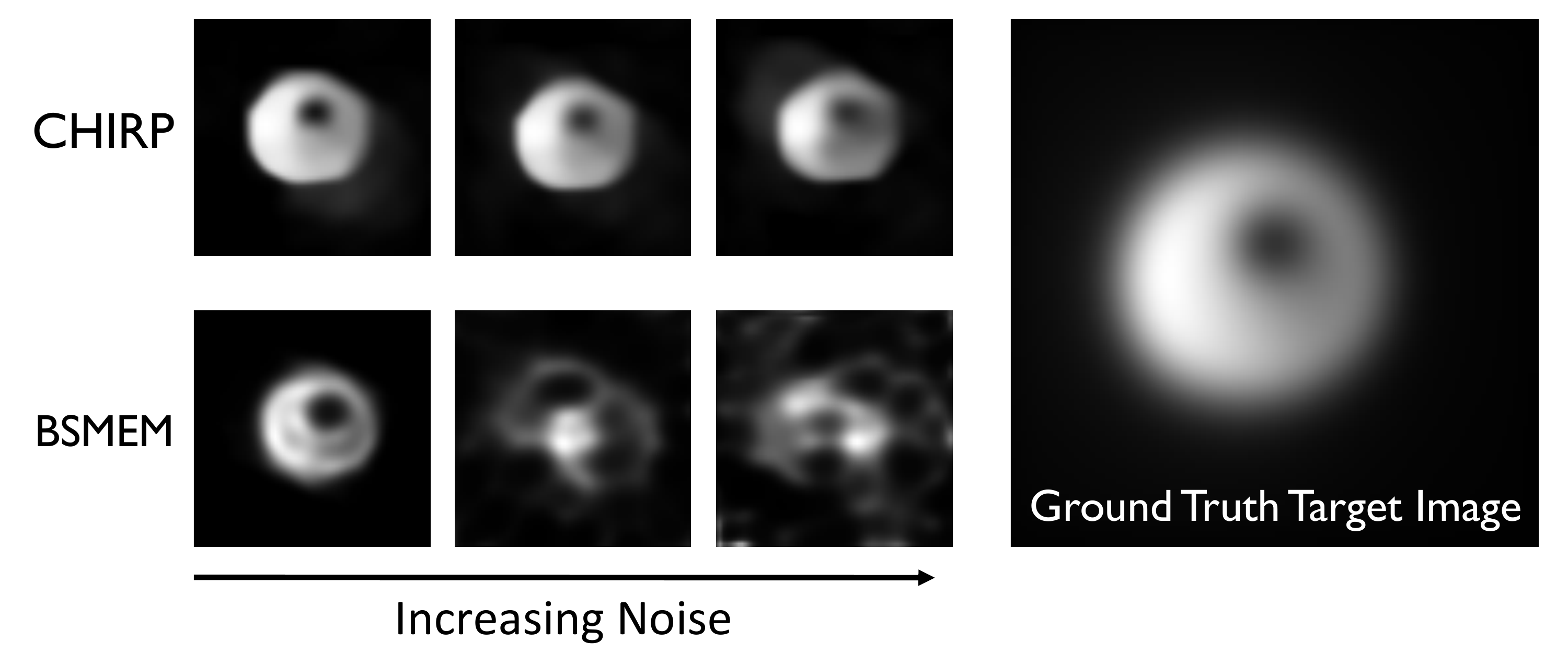}
\caption{(\textbf{a}) Validation of CHIRP on data from the Boston
  University Blazar Group demonstrating superresolution
  \cite{jorstad2005, bouman2016}.  (\textbf{b}) Comparison of CHIRP
  (trained on natural images) and BSMEM on reconstructing a black hole
  model image (courtesy A.\ Broderick) from simulated data.  CHIRP is
  more robust as the signal-to-noise ratio of the simulated data is
  lowered.
  \label{chirp}}
\end{figure}

CHIRP has been validated on a collection of images from the Boston
University Blazar Group \cite{jorstad2005} (Figure~\ref{chirp}).
Under a comparison between imaging algorithms that was unsupervised
(i.e., no user intervention was allowed to fine tune the algorithm for
individual images), CHIRP tended to outperform CLEAN, BSMEM, and
SQUEEZE, especially for datasets with lower signal-to-noise ratios and
for reconstructing images of extended sources.

\section{Summary}

The Event Horizon Telescope has been able to probe Sgr~A*, M87, and
other AGN sources at very high angular resolution, addressing
fundamental physical and astrophysical questions associated with
accreting black holes.  Improvements in sensitivity and baseline
coverage, notably the inclusion of ALMA in 2017, will significantly
increase the capacity of the EHT to produce images.

New imaging techniques are being developed to use robust VLBI
observables to make the most of the relatively sparse baseline
coverage of the EHT.  These techniques have broad applicability beyond
the EHT (e.g., on longer-wavelength VLBI observations).  Validation of
these algorithms on real longer-wavelength VLBI data demonstrate their
potential for reconstructed images with greater resolution and
fidelity than currently provided by CLEAN.  Observers using data from
longer-wavelength VLBI arrays, such as the VLBA and GMVA, may find
these methods useful.

\vspace{6pt} 

\acknowledgments{This work is made possible by grants from the
  National Science Foundation and the Gordon and Betty Moore
  Foundation.}

\bibliographystyle{mdpi}

\renewcommand\bibname{References}

\end{document}